\documentclass[preprint]{JHEP} 

\usepackage{epsfig}                     

\newbox\mybox
\newcommand\fverb{\setbox\mybox=\hbox\bgroup\verb}
\newcommand\fverbdo{\egroup\medskip\noindent\fbox{\unhbox\mybox}\ }
\newcommand\fverbit{\egroup\item[\fbox{\unhbox\mybox}]}

\font\beeg=cmr17 scaled 1600            
\newcommand\init[1]{\setbox\mybox=\hbox{{\beeg #1}~}%
                   \noindent\global\hangindent=\wd\mybox\global\hangafter-2%
                   \sc\smash{\llap {\lower 13.2pt \box\mybox}}}

\newcommand{\lsim}   {\mathrel{\mathop{\kern 0pt \rlap
  {\raise.2ex\hbox{$<$}}}
  \lower.9ex\hbox{\kern-.190em $\sim$}}}
\newcommand{\gsim}   {\mathrel{\mathop{\kern 0pt \rlap
  {\raise.2ex\hbox{$>$}}}
  \lower.9ex\hbox{\kern-.190em $\sim$}}}
\def\be{\begin{equation}}
\def\ee{\end{equation}}
\def\ba{\begin{eqnarray}}
\def\ea{\end{eqnarray}}

\def\ap{\approx}

\def\theta{\vartheta}
\def\epsilon{\varepsilon}

\title{The KARMEN anomaly, light neutralinos and type II supernovae}

\author{M. Kachelrie{\ss}\\
        {\it\small TH Division, CERN, CH--1211 Geneva 23} }

\preprint{CERN-TH 2000-014}

\abstract{
The KARMEN experiment observes a time anomaly in events induced by pion 
decay at rest. This anomaly can be ascribed to the production of a new 
weakly interacting particle $X$ with mass $m_X\ap 34$~MeV. We show that
a recently proposed identification of the $X$ particle with the
lightest neutralino $\chi$ in the frame work of the MSSM with broken
R parity is in contradiction to optical observations of type II supernovae.
}


\begin{document}

{\em Introduction---}
The experiment KARMEN has investigated a variety of neutral
and charged current neutrino interactions finding excellent agreement
between measured cross-sections and the predictions of the standard
model. However, the analysis of the time  distribution of events
induced by neutrinos from $\pi^+$ and $\mu^+$ decays at rest has
revealed an anomaly: the measured distribution for subsequent events
differs substantially from the expected exponential distribution with
a time constant equal 
to the muon lifetime of 2.2~$\mu$s \cite{K1}. As a possible
explanation for this anomaly, the KARMEN collaboration proposed that
their signal is a superposition of a Gaussian distribution centered at
3.4~$\mu$s and the exponential distribution describing muon decay. 
The Gaussian distribution is interpreted as
time-of-flight signature of a hypothetical particle $X$ produced at
the spallation target, passing through 7~m steel shielding and then
decaying in the detector. A maximum likelihood analysis of this
hypothesis showed that the probability that the
Gaussian signal is a statistical fluctuation is only $10^{-4}$.
The best-fit values for the mass of the particle $X$ are $m_X\ap
33.9$~MeV, while the branching ratio BR$(\pi^+\to\mu^++X)$ and the
decay rate $\Gamma_{\rm vis}$ of $X$ into photons and electrons have to
fulfill the relation
BR$(\pi^+\to\mu^++X)\Gamma_{\rm vis}\ap 3\cdot 10^{-11}$s$^{-1}$.
Furthermore, the KARMEN data disfavor two-body decays of the $X$
particles, because no peak at 17~MeV has been seen in the energy
spectra of the anomalous events.  

There have been several theoretical works discussing candidates for the 
new particle $X$ \cite{nu,bo,R1,R2}. Proposed candidates are an active 
or sterile neutrino \cite{nu}, a scalar boson \cite{bo} and a light
neutralino \cite{R1,R2}. While an active neutrino seems to be
excluded by the new experimental limit for BR$(\pi^+\to\mu^++X)$,
a sterile neutrino was found to be, within stringent limits on its
mixing parameters, a viable candidate for the $X$ particle. 
In Ref.~\cite{bo}, a scalar boson with mass $\ap 104$~MeV was proposed
as $X$ particle. Since the energy deposited by the decaying $X$
particle in the calorimeter of the KARMEN experiment is only between 11 
and 35~MeV, additional light scalars have to be invoked to dilute the
energy via cascade decays. Although the model seems to be 
compatible with laboratory constraints, it looks somewhat artificial.
In Ref.~\cite{R1}, the $X$ particle was identified with 
the lightest neutralino $\chi$ in a supersymmetric model with broken R
parity. However, the proposed decay mode of the neutralino
was a two-body decay,
$\chi\to\gamma+\nu_\mu$,  and is therefore now disfavored
by the KARMEN data. Recently, the authors of Ref.~\cite{R2}
reconsidered this proposal. They introduced two R parity violating
operators instead of one, explaining the anomalous pion decay by 
$L_2Q_1D_1^c$ and
the neutralino decay $\chi\to e^+e^-\nu_{\mu,\tau}$ by $L_eL_\mu E_e^c$
or $L_eL_\tau E_e^c$. They found that this scenario is consistent with
accelerator bounds provided that the neutralino is dominantly a bino
and has a lifetime $\tau_\chi = 0.24-100$~s.
Moreover, they showed that regions exist in the MSSM parameter space 
which allow such a bino without excessive fine-tuning.
The same conclusion was obtained already earlier in Ref.~\cite{NP}.

The possibility that KARMEN has observed already over several years a
supersymmetric particle is intriguing. In particular, the scenario 
of Ref.~\cite{R2} with its rather constrained parameter space
would have impact on searches for supersymmetry at accelerators as well
as on searches for dark matter. It is therefore of interest to check
this model against all possible constraints. 
In this brief note, we discuss the influence of neutralino emission on
core collaps supernovae (SN). 
We find that the production of neutralinos with $m_\chi\ap 34$~MeV 
is practically not suppressed in the SN core. The energy
deposited in the SN envelope by decaying neutralinos is for all
allowed squark masses larger than the value allowed by
observations of the light curves of type II supernovae%
\footnote{This argument was first used by S.W.~Falk and D.N.~Schramm
  in Ref.~\cite{fa78} to restrict radiative decays of neutrinos.}. 
We conclude that a light 
neutralino with lifetime $\tau_\chi = 0.24-100$~s is in contradiction 
to observations.

\vskip0.2cm
{\em Neutralino production---}
In the SN core, neutralinos are mainly produced in nucleon-nucleon
bremsstrahlung $NN\to NN\chi\chi$ and in $e^+ e^-$ annihilations. 
The spin-averaged squared matrix element of the latter process 
is for a bino-like neutralino, degenerated selectron masses
$M_{\tilde e}$  and $m_e=0$ \cite{hk} 
\be   \label{M} 
 |\bar M|^2 = \frac{g^{\prime\:4}}{2}  (Y_L^4 + Y_R^4)
 \left\{  \left( \frac{t-M_\chi^2}{t-M_{e}^2} \right)^2
        + \left( \frac{u-M_\chi^2}{u-M_{e}^2} \right)^2
        -  \frac{2s M_\chi^2}{(t-M_e^2)(u-M_{e}^2)} 
 \right\} .  
\ee
Neglecting the Pauli-blocking factors of the neutralinos, the
cross-section times the relative velocity is 
\be
 v\sigma = \frac{1}{(2\pi)^2} \: \frac{1}{8EE'} \int\frac{d^3k}{2\omega}
           \: \frac{d^3k'}{2\omega'} \, |\bar M|^2 
           \: \delta^{(4)} (p+p'-k-k') \,,
\ee
where $p=(E,\vec p)$, $p'=(E',\vec p\,')$ are the four-momenta of the
electron and positron and $k=(\omega,\vec k)$, $k'=(\omega',\vec k')$
of the neutralinos, respectively. 
The emissivity $\epsilon(e^+e^-\to \chi\chi)$ follows as
\be  \label{eps}
 \epsilon(e^+e^-\to \chi\chi) = 
                 \frac{1}{2\pi^4} \int dp dp' d\theta\sin\theta
                 \:\frac{p^2}{e^{(E-\mu)/T}+1} 
                 \: \frac{p^{'2}}{e^{(E'+\mu)/T}+1}
                 \: (E+E') \, v\sigma \,, 
\ee
where $T$ is the temperature and $\mu$ the chemical potential of the
positron-electron plasma.
The integration over the angle $\theta=\angle(\vec p,\vec p\,')$ can be
performed analytically, but results in lengthy expressions. Therefore,
we have preferred to evaluate directly Eq.~(\ref{eps}). In Fig.~\ref{fig}, we
show the ratio $R=\epsilon(x) / \epsilon(0)$ as function of
$x=m_\chi/T$ for different values of the degeneracy parameter
$\eta=\mu/T$. Typical values found in simulations for the temperature
and the electron degeneracy inside the SN core are $T=10-40$~MeV and 
$\eta=10-30$ \cite{bu86}. For an average temperature of $T=20$~MeV,
i.e. $x=1.7$, the neutralino production is
even for the low value $\eta=10$ only reduced by 40\%. A significant
suppression of the neutralino production requires $x\gsim 10$, i.e. 
temperatures well below those believed to exist in SN cores. 
\FIGURE{\epsfig{file=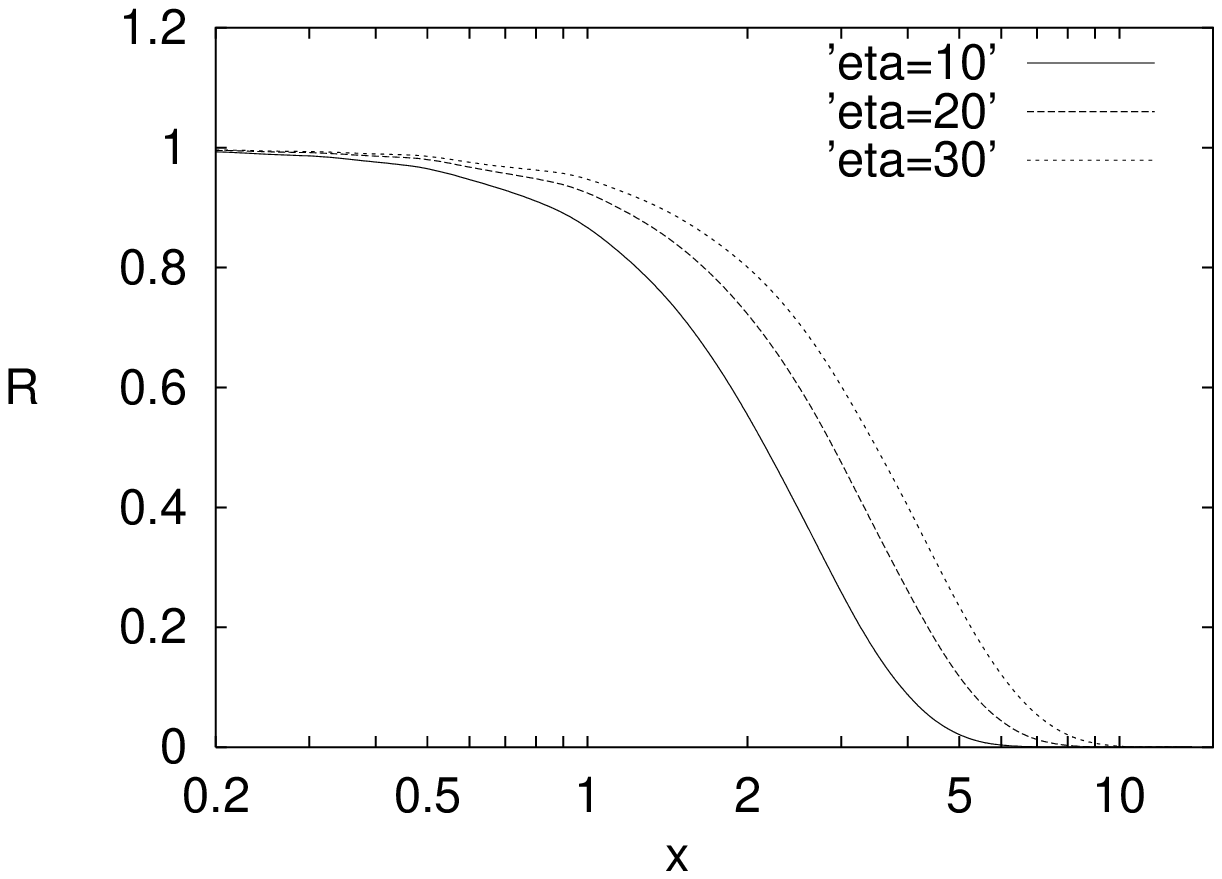,height=7.0cm,width=10.0cm}
        \caption{\label{fig}
         The ratio $R=\epsilon(x) / \epsilon(0)$ as function of
         $x=m_\chi/T$ for $\eta=10, 20$ and 30.}}

It is natural to assume that the bremsstrahlung process $NN\to
NN\chi\chi$ is suppressed roughly by the same factor $f\sim 0.5$. 
This assumption allows us to obtain the emissivity simply from the one
for the related processes with neutrinos \cite{el88}, e.g.,
\be  \label{} 
 \frac{\epsilon(nn\to nn\chi\chi)}{\epsilon(nn\to nn\nu\bar\nu)} \ap 
 f \: \frac{16 m_W^4 \tan^4\theta_W}{M_{\tilde q}^4} 
 \left(\frac{\sum_{i=u,d,s} (Y_{L,i}^2 + Y_{R,i}^2)\Delta q_i}
            {\Delta u -\Delta d - \Delta s}
 \right)^2 \,,
\ee
where $\Delta q_i$ denotes the spin fraction carried by the quark $i$.
Using for $\epsilon(NN\to NN\nu\bar\nu)$ the expression given in
Ref.~\cite{ra} for non-degenerate nucleons and the spin fractions of
Ref.~\cite{spin}, the total energy $E_\chi$ emitted into neutralinos
from the SN in the free-streaming regime is given by
\be
 E_\chi \ap 2.7 \cdot 10^{22}  \: \frac{\rm erg}{\rm g \, s} \: f \: 
            \left(\frac{250 \rm GeV}{M_{\tilde q}}\right)^4 \: 
            \left(\frac{T}{25 \rm MeV}\right)^{5.5} \: 
            \frac{\rho}{2\rho_0} \: 
            \tau_{\rm burst} \, M_{\rm core} \,.  
\ee
With $f=0.5$, $T=25$~MeV, $\rho=2\rho_0\ap 6\cdot 10^{14}$~g/cm$^3$,  
a duration of the neutrino and neutralino burst of 
$\tau_{\rm burst}=10$~s and $M_{\rm core}= 1.5 M_\odot$, we find 
$E_\chi = 4.0\cdot 10^{56} {\rm erg} (250\,{\rm GeV}/M_{\tilde q})^4$ in
the free-streaming regime.

\vskip0.2cm
{\em Energy deposition in the SN envelope---}
An emitted neutralino with lifetime $\tau_\chi$ has the probability
$P\approx 1-\exp(-R/\tau_\chi\gamma)$ to decay inside the SN progenitor 
with radius $R$, where $\gamma=E_\chi/m_\chi$. We will use conservatively 
as radius for the progenitor stars of type-II supernovae the value 
$R\approx 1000$~s \cite{fa}. Thus for all allowed lifetimes
$\tau_\chi=0.24-100$~s a large fraction of the neutralinos decays
inside the envelope, depositing its energy there during the first 10~s
after core collapse.   
There are several possibilities how this energy deposition can be used
to restrict or exclude a light, decaying neutralino. First, the
electromagnetic displaying of the SN which is expected to start 3 hours
after the neutrino burst, when the shock wave reaches the photosphere,  
should start much earlier. Second, model calculation for 
SN1987A find that the total energy of the shock is 
$E_{\rm sh}\ap 3\cdot 10^{51}$~erg. This number can be used as upper
limit for the energy $E_{\rm en}$ released by the neutralinos in
the envelope.

Assuming that the neutrino carries away one third of the energy, 
neutralinos deposit the energy $E_{\rm en} = \frac{2}{3} E_\chi P$
in the SN envelope. With $P>0.63$, we can exclude those parts of the 
parameter space of the model~\cite{R2} which give rise to 
$E_\chi\gsim 7\cdot 10^{51}$~erg. Thus, squark masses smaller
than $\ap 4$~TeV are excluded in the free-streaming regime.

For very light squarks, nucleon-neutralino interactions become 
strong enough so that neutralinos are efficiently trapped and $E_\chi$ 
falls below $7\cdot 10^{51}$~erg. 
The results of Ref.~\cite{el88} suggest, that this
possibility is already excluded by direct accelerator searches. 
Nevertheless, we will reconsider below the trapping regime for a bino
like neutralino.

\vskip0.2cm
{\em Neutralino opacity---}
In the calculation of the free-mean path of the neutralino in the SN
core, we should take into account the thermal distribution functions
of the electrons and nucleons, respectively. A detailed calculation
performed in Ref.~\cite{la93} for massless photinos showed that the thermal 
cross-section of $\chi e^-\to\chi e^-$ can be rather well approximated
by the vacuum one. Moreover, the thermal effects can be factored out
in neutralino-nucleon scattering, when recoil effects are neglected.
 
If selectrons are not considerable lighter than the lightest squarks, the
dominant opacity source for neutralinos is scattering on nucleons.
We estimate the free-mean path $l_\chi$ of the neutralino as 
\be
 l_\chi^{-1} (\chi p) \ap 4.2 \cdot 10^{-6} {\rm cm}^{-1}  
                      \left(\frac{250{\rm GeV}}{M_{\tilde q}}\right)^4 
                      \: \frac{(1-Y_n)\rho}{2\rho_0} \,,
\ee
\be
 l_\chi^{-1} (\chi n) \ap 1.0 \cdot 10^{-6} {\rm cm}^{-1} 
                      \left(\frac{250{\rm GeV}}{M_{\tilde q}}\right)^4 
                      \:\frac{Y_n\,\rho}{2\rho_0} \,,
\ee
where we used as average energy of the neutralino 
$E_\chi \ap 3T \ap 75$~MeV and a thermal suppression factor $0.7$
(cf. \cite{la93}).
Requiring $l_\chi^{-1}R=10$ as trapping criteria with $\rho=2\rho_0$,
$Y_n=0.8$,  
and $R=10$~km as size of the core, we find $M_{\tilde q}\ap 160$~GeV
as borderline between the diffusion and the free-streaming regime.

In the diffusion regime, we estimate the neutralino luminosity 
${\cal L}_\chi$ assuming blackbody surface emission,
${\cal L}_\chi= (\pi^3/15) \: R_\chi^2 T_\chi^4$,
from a $\chi$-sphere with radius $R_\chi$. The position of this
sphere is calculated from the optical depth $\tau$,
$\tau=\int_{R_\chi}^\infty dr\: l^{-1}_\chi(r)=2/3$, where we use as
density profile $\rho(r)=\rho_R(R/r)^n$, as temperature profile 
$T(r)=T_R(\rho(r)/\rho_R)^{1/3}$ and $Y_n=0.5$ \cite{ra}. 
We choose density and temperature at the edge of the core as
$T_R=10$~MeV and $\rho_R=10^{14}$g/cm$^3$, respectively. 
The exponent $n$ of the profiles is rather model dependent, $n\sim3-7$. 
Using the scaling relation 
\be
 L_\chi(r)=L_\chi(R) \left(\frac{R}{r}\right)^{\frac{4n}{3}-2} 
\ee
and $n=5$,
the neutralino luminosity is small enough for $R_\chi\gsim 2.8 R$.
This is consistent with $\tau(R_\chi)=2/3$ only if the squark masses
would be below 20~GeV.

\vskip0.2cm
{\em Summary---}
The production of neutralinos with the mass $m_\chi\ap 34$~MeV is
practically unsuppressed in a SN core. The upper bound on the sfermion
masses in the model of Ref.~\cite{R2}, $M_{\tilde f}\lsim 1$~TeV,
together with the lower limit $M_{\tilde f}\gsim 100$~GeV from LEPII
leaves no room for a decaying neutralino with the required lifetime 
$\tau_\chi = 0.24-100$~s:
For all allowed values of $M_{\tilde f}$, the energy deposited in the
SN envelope by decaying neutralinos is in contradiction to the optical
observations of type II supernovae.

\vskip0.6cm
{\em Note added:\/}
The preprint hep-ph/9912465 by I.~Goldman, R.~Mohapatra and
S.~Nussinov discusses also the
consistency of a light, neutral, decaying fermion $n^0$ with SN 1987A. 
Its discussion is model-independent, using however the assumption that
the $n^0$-nucleon cross-section is at least a factor of 10 smaller
than the neutrino-nucleon cross-section. This assumption does not hold
necessarily in the case of the neutralino.

\end{document}